\documentstyle[preprint,aps]{revtex}
\begin{document}
\draft
\title{
Classical and Quantal Irregular Scatterings \\
with Complex Target
}
\author{
Taksu Cheon
}
\address{
Department of Physics, Hosei University
Chiyoda-ku Fujimi, Tokyo 102, Japan
}
\date{June 1, 1995}
\maketitle

\begin{abstract}

One-dimensional scattering by a target with two internal 
degrees of freedom is investigated.  The damping of resonance peaks 
and the associated appearance of the fluctuating background in the 
quantum inelastic scattering amplitudes are found.  Examination of the 
analog classical system reveals a disorderly reaction function, which is 
then related to the quantum amplitude through a semiclassical 
argument.

\end{abstract}
\narrowtext

\newpage

\section{INTRODUCTION}

	Already three decades ago, Ericson [1] recognized that the
background fluctuations of the quantum scattering amplitudes carry as
much physical information as the structure of resonance peaks.  It is
the recent advent of the theory of the non-integrable scatterings that
has begun to unveil the intimate connection between the irregular
scattering and the Ericson fluctuation, or the fluctuation of the
elastic scattering amplitude when the incident bombarding energy is
varied [2-4].  However, the continuous variation of the incident energy is
hard to materialize experimentally in particle and nuclear physics.
More often, the scattering experiments are performed with fixed energy
accelerators, but with particle detectors that allow the detection of
the scattering observables at various outgoing energies.  There, one
typically observes several large peaks corresponding to simple
excitation modes and the fluctuating backgrounds in between.  It is
well known that the nature of the target dynamics determines the
characteristics of these resonance peaks and the background
fluctuations.  Namely, when the mixing interactions among target
states are present, one usually finds damping of the peaks and the
resulting appearance of "noisy" background.  Bearing the Ericson
fluctuation in mind, it is natural to speculate on the possible
relevance of the behavior of the inelastic amplitudes to the chaotic
aspects of the system.  In this note, we study a model system in which
a one-dimensional particle is scattered by a target which has two
dynamical degrees of freedom.  In the quantum scattering, we find that
the damping of the peaks indeed occurs when the target goes through
complex motion.  The study of the classical analog system reveals the
existence of a novel type of irregular scattering in which the
location of the maxima and minima of the reaction function becomes
disorderly.  We discuss a possible way to connect the behavior of
classical and quantum systems in scattering.

\section{MODEL AND TARGET DYNAMICS}      

 We consider a projectile moving in one-dimensional space specified by
the coordinate and its conjugate momentum $(x, p)$.  The particle is
scattered by a system (or the target) which is described by the two
internal degrees of freedom specified by two sets of "angle" and
conjugate "angular momentum" $(\vartheta, I)$ and $(\varphi, J)$.
We assume that the
target system is 2$\pi$-periodic, and treat the variables 
$\vartheta$ and $\varphi$ as in the range of [-$\pi, \pi$]. 
The dynamics of the system of a projectile and
a target is governed by the Hamiltonian
\begin{eqnarray}
\label{1}
H={{p^2} \over {2m}}+\left[ {{{I^2} \over {2M}}
+{{J^2} \over {2N}}+U(\vartheta ,\varphi )} \right]
+V(x,\vartheta ,\varphi )
\end{eqnarray}
where the terms inside the bracket, which we call target Hamiltonian $H_I$, 
describe the motion of the target with the internal coupling interaction
\begin{eqnarray}
\label{2}
U(\vartheta ,\varphi) 
= {1 \over {2\pi }}(1-3 \cos \vartheta+2 \cos 2\vartheta)
  {1 \over {2\pi }}(1-3 \cos \varphi+2 \cos 2\varphi)
\end{eqnarray}
The particle-target interaction $V$ is given by
\begin{eqnarray}
\label{3}
V(x,\vartheta ,\varphi )=\lambda \cdot {1 \over {\sqrt \pi }}\exp (-x^2)\
 \cos 2\vartheta
\end{eqnarray}

Here $s$ and $\lambda$ are the parameters 
which control the overall strength of
target-internal and particle-target coupling interactions
respectively.  Throughout this work, we adopt such unit as to make 
$\hbar$ = 1.  The masses are fixed to be $m$ = 1, $M$ = $\sqrt 2$
and $N$ = $\sqrt 3$.

	We denote the eigenstates and eigenvalues of the target
Hamiltonian $H_I$ as $\{ \left| {\left. \alpha \right)} \right. \}$ and
$\{ \varepsilon_\alpha \}$.  When the coupling is absent
($s$ = 0), the target state is
reduced to the "free rotation" in both angles 
$\vartheta$ and $\varphi$, and is
specified by two integer quantum numbers $m$ and $n$, namely, $\left|
{\left. \alpha \right)} \right.$
 = $\left| {\left. {m,n} \right)} \right.$ .  The effect of the mixing
interaction $U$ on the global properties of the target states is
conveniently characterized by the nearest neighbor spacing
distribution of eigenvalues $P(s)$ [5,6].  
In Figs. I (a)-(c), we plot $P(s)$
for the coupling strengths $s$ = 0, 1 and 10 along with the Poisson
distribution (dashed line) and the Wigner distribution (solid line)
which are usually associated with regular and chaotic dynamics
respectively.  There are 200 lowest levels 
(between $\varepsilon$ = 0 -- 75)
included in the calculation.  We clearly observe the gradual approach
to the chaotic spectra with the increasing value of $s$.  To strengthen
the point, we also plot in Figs. I (d)-(f) the Poincare section 
$(\vartheta,I)$, with the section of $\varphi$ = 0 
of the target Hamiltonian $H_I$, which is
obtained from the solution of classical equations of motion using the
same parameter sets as in Figs. I (a)-(c).  Since the target
Hamiltonian does not possess the mechanical similarity, the phase
space structure depends on the target energy.  In practice, however,
this dependence is found to be very mild in the energy range we are
interested in.  Here, we take the target energy to be $\varepsilon$ = 10.
The transition from regular to chaotic target motion is better visible in
these Figures.

\section{QUANTUM SCATTERING}

	The quantum scattering is most efficiently described with the
transition matrix formalism [7].  The quantum state of the system is
specified by the direct product of the particle state $\left| p
\right\rangle $ and the target eigenstate $\left| {\left. \alpha
\right)} \right.$ .  We refer to the quantum states with internal wave
function $\left| {\left. \alpha \right)} \right.$
as belonging to "channel" $\alpha$.  The inelastic transition between
channel $\alpha$ and $\beta$ occurs through the operation of the interaction
$V_{\beta \alpha}(p',p)$ 
$\equiv \left\langle {p} \right|\left. {\left( \beta \right.} 
\right|V\left| {\left. \alpha \right)} \right.\left| p
\right\rangle $.  
The the transition matrix (T-matrix) from channel $\alpha$
to $\beta$ is obtained from the coupled-channel Lippmann-Schwinger equation
\begin{eqnarray}
\label{4}
T_{\beta \alpha }(p_\beta ,p_\alpha ;\omega )=V_{\beta \alpha }(p_\beta 
,p_\alpha )+\sum\limits_\gamma  {\int {{{dp} \over {2\pi }}V_{\beta 
\gamma }(p_\beta ,p)G_\gamma ^{(0)}(p;\omega )\ T_{\gamma \alpha 
}(p,p_\alpha ;\omega )}}
\end{eqnarray}
where the free particle propagator in the channel $\alpha$ is given by 
$G_\alpha^{(0)}(p;\omega )$
 = $(\omega -p^2/ 2m-\varepsilon _\alpha +i0)^{-1}$ .  The scattering
matrix (S-matrix) between the initial channel $\alpha$ with particle incident
momentum (meaning the asymptotic initial momentum) $p_\alpha$ and the final
channel $\beta$ with particle outgoing momentum $p_\beta$ is given by

\begin{eqnarray}
\label{5}
S_{\beta \alpha }(p_\beta ,p_\alpha ;\omega )\ 
=\ \delta _{\beta \alpha }\ +\ {m \over {i\sqrt {\left| {p_\beta p_\alpha } 
\right|}}}\cdot T_{\beta \alpha }(p_\beta ,p_\alpha ;\omega )
\end{eqnarray}
where the on-shell condition $\omega$ = $p_\beta ^2/ 2m$ +$\varepsilon_\beta$ 
= $p_\alpha^2/ 2m$ +$\varepsilon_\alpha$
is imposed on the momenta $p_\alpha$ and $p_\beta$.  All the scattering
observables are obtainable from the S-matrix elements.

	We look at the behavior of the S-matrix elements with the
change of the final state $\beta$, while keeping 
the initial state $\alpha$ to be the ground state and 
also keeping the incident energy $p_\alpha ^2/2m$ constant.  
In Fig. II, the absolute value of the S-matrix is
plotted as a function of energy difference between the final and
initial target states $(\varepsilon_\beta- \varepsilon_\alpha)$.
The incident momentum is set to be $p_\alpha$ = 6.5 
(or $p_\alpha ^2/ 2m$ = 21.125) which is capable of exciting
about 60 target levels.  The overall strength of the particle-target
interaction is chosen to be $\lambda$ = 50.  This choice makes the strength of
the interaction comparable to the kinetic energy of the incoming
particle $p_\alpha ^2/ 2m$, and gives the sufficient strength to the
inelastic amplitudes.  The Figs. II (a) to (c) show the results of the
scattering by the target states with different mixing properties.
Each graph corresponds to the one with same alphabet of Fig. I,
namely; (a) no mixing with $s$ = 0, (b) weak chaotic mixing with $s$ = 1
and (c) strong chaotic mixing with $s$ = 10.  Only scattering amplitudes
to the positive value of pb is shown.  The backward scattering
amplitudes (found to be generally small with the model interaction $V$)
show similar characteristics.  Because of the selection rule of the
matrix elements of the interaction $V$, the scattering from integrable
target shows a simple pattern of exciting very few states which form
the sharp peaks in the Fig. II (a).  With the complex target states
(Figs. II (b) and (c)), the peaks of the scattering amplitude begin to
loose their strength to various other states with strength seemingly
at random.  To understand this behavior, one needs to look at the
matrix element $\left. {\left( \beta \right.} \right|T\left|
{\left. \alpha \right)} \right.$
 = $\sum {_{n,m,n,m}}\left( \beta  \right.\left| {\left. {n,m} \right)} 
\right.\left. {\left( {n,m\ } \right.} \right|T\left| {\left. {n,m} \right)} 
\right.\left( {n,m} \right.\left| {\left. \alpha  \right)} \right.$
When the coupling s increases, each target state $\left| {\left. \alpha  
\right)} \right.$
 becomes a complex mixture of unperturbed states $\left| {\left. {n,m}
\right)} \right.$ and the summation by n and m will cause generally
incoherent interferences, resulting in the white-noise like behavior
of $\left. {\left( \beta \right.} \right|T\left| {\left. \alpha
\right)} \right.$ .  This mechanism by itself is frequently
encountered in atomic and nuclear physics as the partitioning of
excitation strength of simple states to more complex configurations.
We now ask a question: Is there any the property of underlying
classical dynamics that is relevant to the quantum phenomenon of the
spreading of resonance peaks?

\section{CLASSICAL SCATTERING}

	It is nowadays widely recognized 6 that the deeper
understanding of quantum phenomena is achieved by the examination of
the classical trajectories 
$\{ x(t), p(t), \vartheta(t), I(t), \varphi(t), J(t) \} $ which
is the solution of the classical equations of motion obtained from the
Hamiltonian eq. (1).  The scattering process is described by the
reaction functions,8 or the functional relation between the "outgoing"
variables such as 
$p_{out} = p(T)$ and $x_{out} = x(T)-p(T)T/m$ (where $T$ is
sufficiently lager than the reaction time) and the "incoming"
variables $p_{int} = p(0)$, $x_{in} = x(0)$, 
$I_{in} = I(0)$, $J_{in} = J(0)$ etc..
 Here, we look at the particular reaction function
\begin{eqnarray}
\label{6}
p_{out} = p_{out}(x_{in})
\end{eqnarray}
while keeping all other incoming variables constant.  In Fig. III we
display the typical case.  The initial momentum is taken to be $p_{in}$ =
6.5 as before.  The initial values of the other dynamical variables
are taken to be $\vartheta_{in}$ = $\varphi_{in}$ = 0 
and $I_{in}$ = $J_{in}$ = 1.0 in this example.
The choice of the strength parameter $s$ in each of Figs. III (a)-(c)
corresponds to the same alphabet in previous figures.  In all cases,
the reaction functions display singular structure (dotty areas) which
is the reflection of the non-integrability of the model system as a
whole.  The irregular structure in case of no internal coupling
(Fig. III (a)) is essentially identical to the one previously
encountered in the study of Blummel and Smilansky [2].  It is related
to the instability of the "trapping" orbits.  It is worth noting that
the irregular area appears in regular interval in xin reflecting the
periodicity (or quasi- periodicity) of the target motion.  When the
target-internal coupling is turned on (Figs. III (b) and (c)), the
nature of the reaction function seems to acquire qualitatively
different irregularity.  Not only the trapping areas occur at
irregular locations but the smooth areas get disorderly shape
deformation.  It is understood as the result of the chaotic motion of
the target which makes the value of the target variables practically
unpredictable at the time of the particle hitting the target.  Thus,
one can conclude that the disorderly reaction function is a generic
property of the scatterings by complex targets.

	Semiclassical consideration should relate the classical and
quantum scatterings.  We extend the argument by Smilansky [8] to the
current model with two target degrees of freedom.  We prepare an
ensemble of initial conditions of the target motion with constant
energy.  This can be done by bringing 
various $\vartheta_{in}$, $I_{in}$, $\varphi_{in}$ and $J_{in}$
while keeping internal energy constant.  When the motion of the target
is ergodic or quasi-periodic, another way of achieving this ensemble
is preparing different xin distributed uniformly (with density one)
over certain region beyond the interaction range.  This is understood
by noticing that advancing the projectile position xin is equivalent
to letting target variables move backward in time.  Then, the outgoing
value pout will be distributed with a probability density $P(p_{out})$
which is given by

\begin{eqnarray}
\label{7}
P(p_{out}) dp_{out} = \sum\limits_s {\left[ {dx_{in}} \right]_{(s)}}
\end{eqnarray}

where the summation ($s$) is over the initial values for
$x_{in}$ satisfying $p_{out} = p_{out}(x_{in})$.  
Semiclassically, the square of the scattering
matrix should approach the classical probability density.  Therefore,
we have the classical approximation to the S-matrix
\begin{eqnarray} 
\label{8} 
\left| {S^{cl}(p_{out},p_{in})} \right|\ =\
\sqrt {P(p_{out})}=\sqrt {\sum\limits_s {\left[ {{{dp_{out}} \over
{dx_{in}}}} \right]_{(s)}^{\ - 1}}}\ 
\end{eqnarray}
This expression shows that the classical S-matrix diverges at the
extrema of the reaction function $p_{out}(x_{in})$.  Presuming that the
quantum corrections amend the divergence and make it a finite peak
instead, we can relate our findings in classical and quantum
scatterings: Mostly smooth reaction function with several repeating
minima and maxima will result in the simple quantum amplitude with few
distinct peaks, while the reaction function with disorderly located
extrema should make the fluctuating background in the scattering
amplitude.  To demonstrate these arguments, we construct the classical
S-matrix, eq. (8) from the reaction function shown in Fig. III.  As in
the full quantum treatment, $|S^{cl}|$ is plotted in Fig. IV as the
function of energy transfer 
$\varepsilon_\beta-\varepsilon_\alpha$ = $(p_{in}^2-p_{out}^2 )/2m$.  Each
of the graph (a)-(c) corresponds to the same alphabet of both Figs. II
and III.  Although no quantitative agreement is observed (nor is it
expected at the value of $\hbar$ = 1), one notices remarkable qualitative
similarities between classical and quantal results.  Specifically, the
key feature of our interest is already present in classical S-matrix;
that is, the partitioning of the strength of large peaks to more
numerous smaller peaks as the motion of the target becomes chaotic.
It is possible to improve the classical approximation of the S-matrix
by incorporating the action integral and Maslov index to include
quantum phases in the manner of Miller [9]:

\begin{eqnarray}
\label{9}
S^{sc}(p_{out},p_{in})\ =\ \sum\limits_s {\sqrt {\left[ {{{dp_{out}}
\over {dx_{in}}}} \right]_{(s)}^{\ -1}}\cdot \exp \left\{
{iS(p_{out},p_{in})_{(s)}-i{{\nu _{(s)}\pi } \over 2}} \right\}}\
\end{eqnarray}

The effect of the inclusion of the phase factor can be seen in the
Fig. V.  Since the expression eq. (9) does not guarantee the unitarity
of the S-matrix, we have numerically renormalized the results to make
the total transition probability unity.  With this admittedly ad hoc
procedure, we can still observe the general improvement toward the
full quantum results, while leaving the essential points made with
classical approximation intact.

\section{CONCLUSION}

	We have studied the phenomenon of damping (or the spreading)
of the peaks in the scattering off the complex target from a
semiclassical point of view.

\acknowledgements
We thank Drs. T. Shigehara and N. Yoshinaga for their 
collaboration.

\begin{figure}
\caption{
Dynamics of  the target Hamiltonian $H_I$. Quantum level statistics $P(s)$ 
and the classical Poincare sections $\varphi$ = 0 
at the energy $\varepsilon$ = 10
}
\end{figure}

\begin{figure}
\caption{
The absolute value of inelastic S-matrix at $p_\alpha$ = 6.5 
with $\lambda$ = 50 as a function of excitation energy.
}
\end{figure}

\begin{figure}
\caption{
Classical reaction function $p_{out}(x_{in})$ at $p_{in}$ = 6.5.
\ \ \ \ \ \ \ \ \ \ \ \ \ \ \ \ \ \ \ \ \ \ \ \ 
\ \ \ \ \ \ \ \ \ \ \ \ \ \ \ \ \ \ \ \ \ \ \ \ 
\ \ \ \ \ \ \ \ \ \ \ \ \ \ \ \ \ \ \ \ \ \ \ \ 
}
\end{figure}

\begin{figure}
\caption{
Classical approximation to the quantum S-matrix, $S^{cl}$.
\ \ \ \ \ \ \ \ \ \ \ \ \ \ \ \ \ \ \ \ \ \ \ \ 
\ \ \ \ \ \ \ \ \ \ \ \ \ \ \ \ \ \ \ \ \ \ \ \ 
}
\end{figure}

\begin{figure}
\caption{
Semiclassical approximation to the quantum S-matrix, $S^{sc}$.
\ \ \ \ \ \ \ \ \ \ \ \ \ \ \ \ \ \ \ \ \ \ \ \ 
\ \ \ \ \ \ \ \ \ \ \ \ \ \ \ \ \ \ \ \ \ \ \ \ 
}
\end{figure}


\begin{thebibliography}{10}

\bibitem{ER63}
T. Ericson, Ann. Phys. (NY). {\bf 23}, 390 (1963).

\bibitem{BS88}
R. Blumel, U. Smilansky, Phys. Rev. Lett. {\bf 60}, 477 (1988); 
{\em ibid.}, {\bf 64}, 241 (1990).

\bibitem{RB91}
A. Rapisarda and M. Baldo, Phys. Rev. Lett. {\bf 66}, 2581 (1991).

\bibitem{SE93}
P. Seba, Phys. Rev. {\bf E47}, 3870 (1993).

\bibitem{BO91}
O. Bohigas, in Chaos and Quantum Physics, eds. M.-J. Giannoni 
and A. Voros, Les Houches Lectures, Session LII 
(North-Holland, Amsterdam, 1991).

\bibitem{GU90}
M. C. Gutzwiller, Chaos in Classical and Quantum Mechanics 
(Springer, Berlin, 1990).

\bibitem{GW64}
See the standard textbook; M. L. Goldberger and K. M. Watson, 
Collision Theory, (Wiley, New York, 1964)

\bibitem{SM91}
U. Smilansky, in Chaos and Quantum Physics, eds. 
M. -J. Giannoni and A. Voros, Les 
Houches Lectures, Session LII (North Holland, Amsterdam, 1991).

\bibitem{MI74}
W. H. Miller, Adv. Chem. Phys. 25, {\bf 69} (1974).

\end{thebibliography}
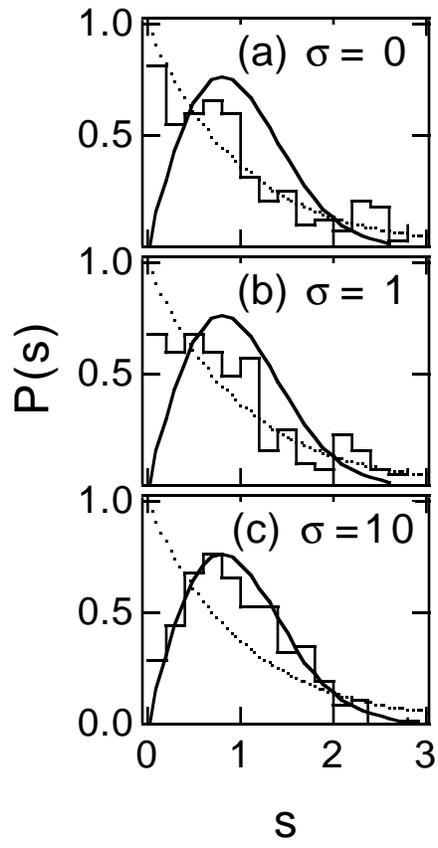
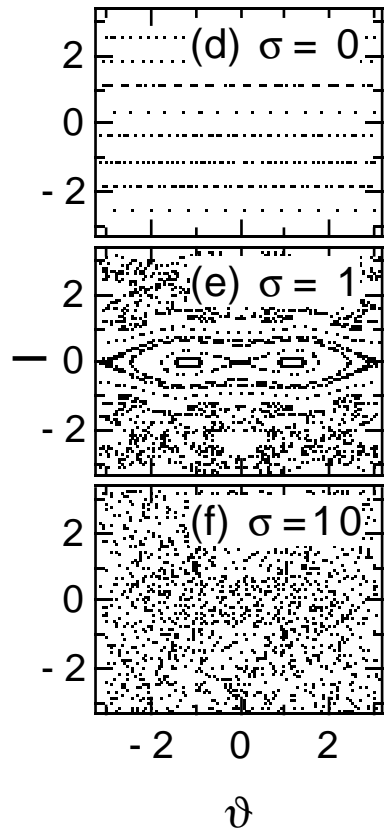
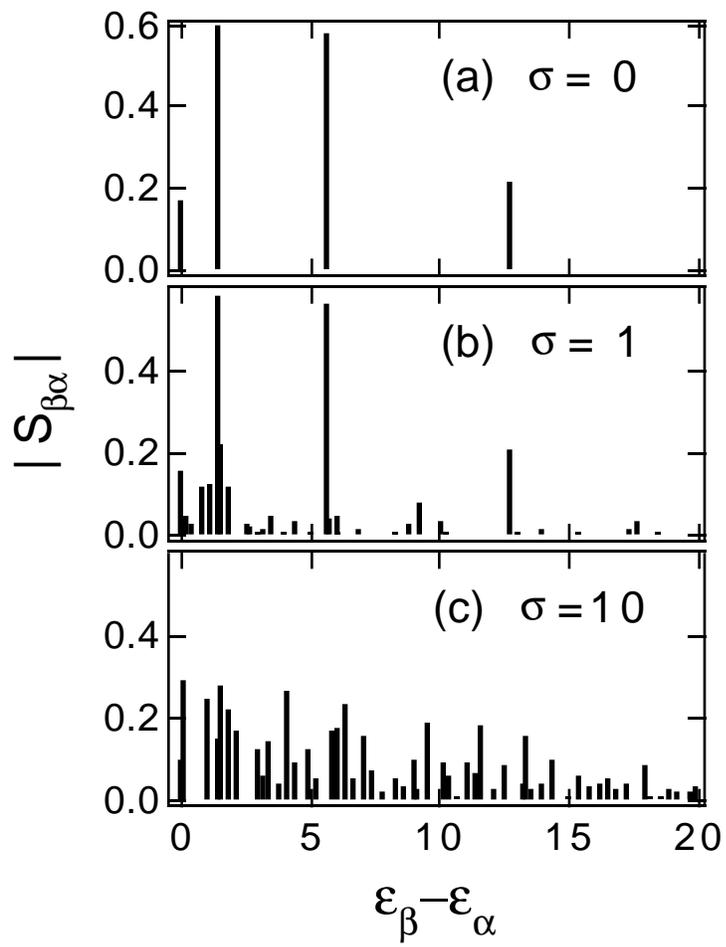
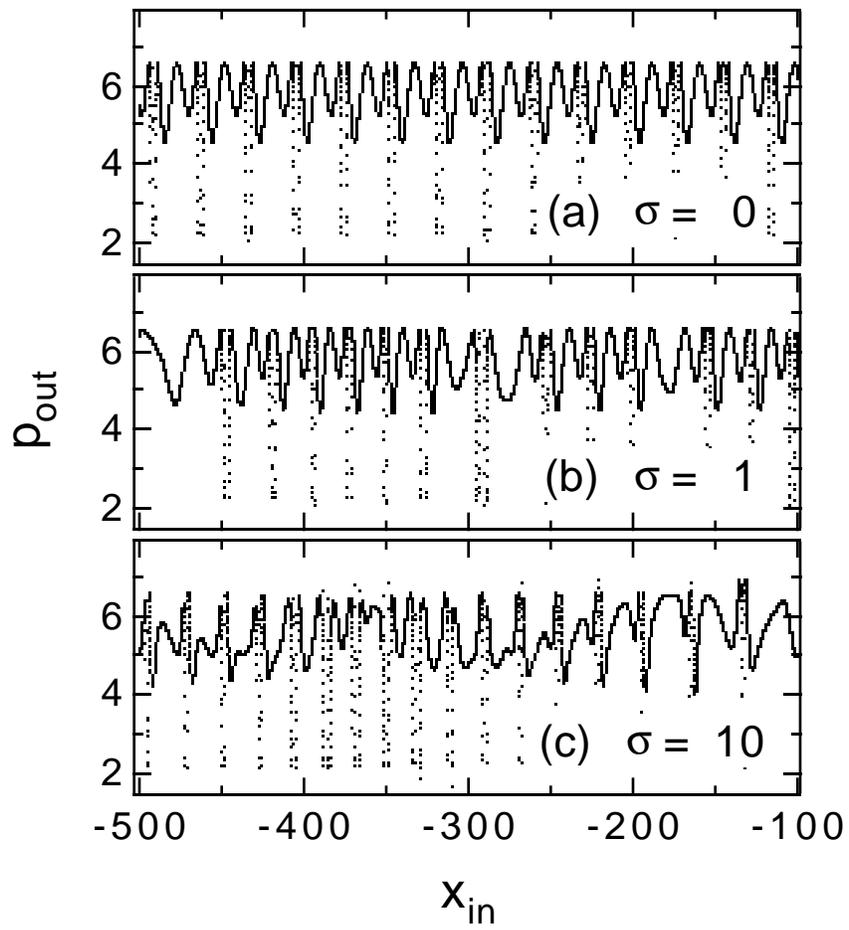
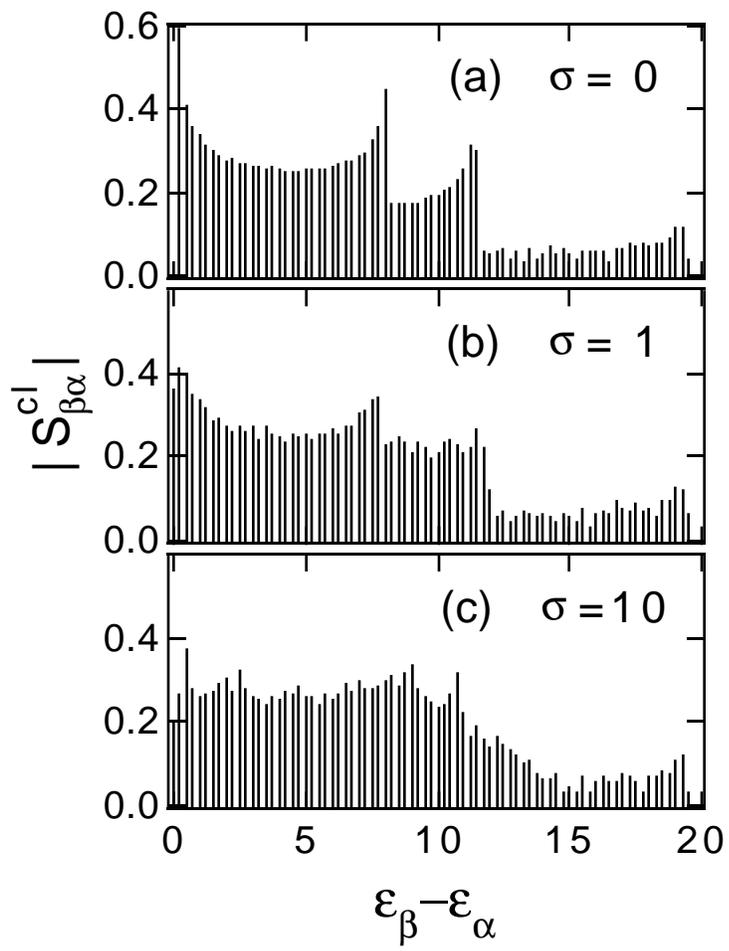
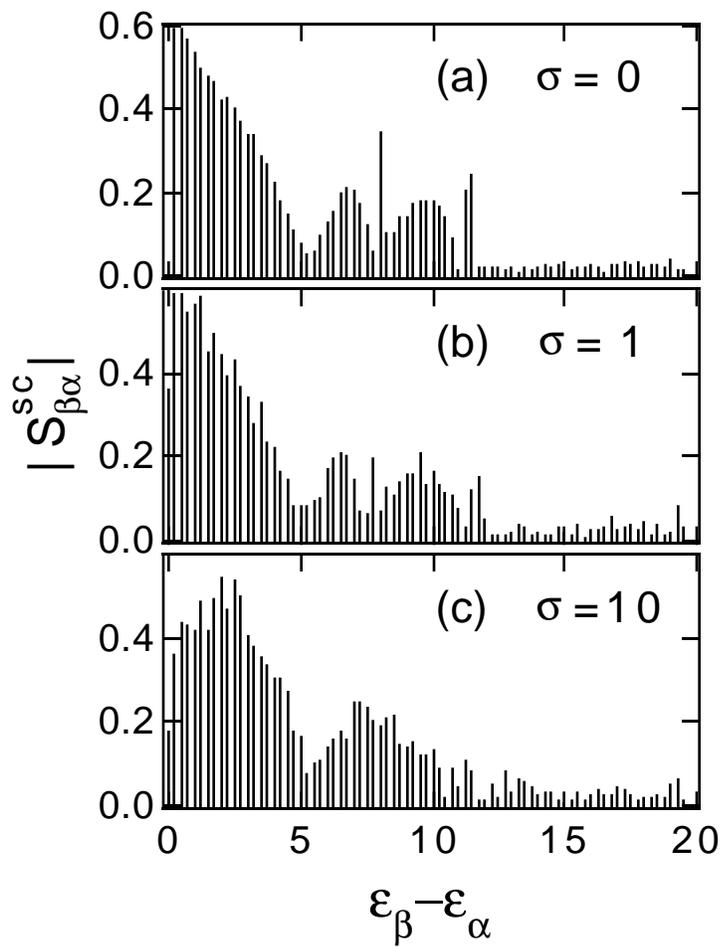
\end{document}